# Unraveling Anisotropic Hybridizations of Solid-state Electrolyte Nano-films in Li-ion Batteries

Yuanjie Ning[1], Wenjun Wu[1], Liang Dai[1], Shuo Sun[1], Zhigang Zeng[1], Dengsong Zhang[2], Mark B. H. Breese[3,4], Chuanbing Cai[1], Chi Sin Tang[3,*], Xinmao Yin[1,*]

[1]Shanghai Key Laboratory of High Temperature Superconductors, Institute for Quantum Science and Technology, Department of Physics, Shanghai University, Shanghai 200444, China

[2]International Joint Laboratory of Catalytic Chemistry, College of Sciences, Shanghai University, Shanghai 200444, China

[3]Singapore Synchrotron Light Source (SSLS), National University of Singapore, Singapore 117603

[4]Department of Physics, Faculty of Science, National University of Singapore, Singapore 117542, Singapore

*To whom correspondence should be addressed: slscst@nus.edu.sg (C.S.T.); yinxinmao@shu.edu.cn (X.Y.)

**ABSTRACT**: $Li_2WO_4$ (LWO) is recognized for its potential as a solid-state electrolyte and it has demonstrated the ability to enhance the electrochemical performance of $LiCoO_2$ (LCO) cathodes in Li-ion batteries. However, prior investigations into LWO have predominantly involved polycrystalline structures, thereby lacking a comprehensive understanding of its behavior when interfaced with single crystal systems, particularly those intricately connected to LCO. In this study, we employ pulsed laser deposition (PLD) to epitaxially synthesize LWO nano-films on LCO layers with different orientations. Based on a series of high-resolution synchrotron-based techniques including X-ray absorption spectroscopy (XAS) and X-ray photoemission spectroscopy (XPS), the electronic structure of LWO is carefully scrutinized where a higher main energy level of $W5d(e_g)$-$O2p$ orbitals hybridization in LWO/LCO(104) as compared to LWO/LCO(003) has been observed. This experimental finding is further validated by a comprehensive set of density of states calculations. Furthermore, detailed polarized XAS characterization unveils distinct anisotropy between the two oriented LWO configurations. This comprehensive scientific investigation, harnessing the capabilities of synchrotron-based techniques, provides invaluable insights for future studies, offering guidance for the optimized utilization of LWO as a solid-state electrolyte or modification layer for LCO cathodes in high-powered Li-ion batteries.

## 1. Introduction

LiCoO$_2$ (LCO) has garnered considerable attention as a cathode material for lithium ion batteries (LIBs) owing to its commendable rate performance, exception reliability and high theoretical capacity[1–4]. Despite these merits, challenges such as capacity decay and voltage fading emerge in LCO cathodes operating at high voltages attributable to interfacial reactions and structural damages[5]. To address these issues, surface coating has been proposed as a solution, wherein a thin-film layer of transition metal oxide between the cathode and electrolyte to enhance the LCO cathode performance[6–10].

Amongst a range of promising coating materials, Li$_2$WO$_4$ (LWO) stands out as a potential solid Li-electrolyte with high ionic conductivity but limited electronic conductivity[11]. Utilised as a coating for LCO cathodes[12–14], LWO serves a dual-purpose. On the one hand, it shields the LCO surface from electrolyte reactions and inhibits the diffusion of Co and O as a protective layer. On the other hand, LWO aids in improving the diffusion of lithium ions, consequently reducing interfacial resistance [12]. While limited instances of amorphous LWO have been reported[6,11,14], research on single crystal LWO film remains sparse. Notably, LCO thin films can be epitaxially grown on SrTiO$_3$ (STO) using pulsed laser deposition (PLD) [15–17], and recent studies have demonstrated the formation of single crystal LWO with controlled orientations on LCO seed layers through spontaneous lithiation [18], wherein the orientation of the STO substrate determines that of both LCO and LWO layers.

Here, we report a markedly distinct electronic structure of single-crystal LWO islands that arise when they are epitaxially grown on LCO layers with lattice different orientations via spontaneous lithiation. A comprehensive synchrotron-based X-ray absorption spectroscopy (XAS) characterization complemented by high-resolution X-ray photoelectron spectroscopy (XPS) and X-ray diffraction (XRD) reveal distinct anisotropic effect and orbital hybridization between the O2p and W5d(e$_g$) orbitals belonging to LWO as a result of the crystal orientation of the LCO bottom layer (Figure 1). This anisotropic structural and orbital properties may be attributed to the WO$_6$ octahedral distortion that arise due to the diffusion of orientation-dependent Li and the unique growth modes of the LWO film based on different LCO crystal orientations. The detailed investigation of the structural and orbital properties that arise at the LWO/LCO interface plays a critical role that enables a deeper mechanistic understanding of how different substrate orientations can dictate and tune the electronic structures of single-crystal LWO systems.

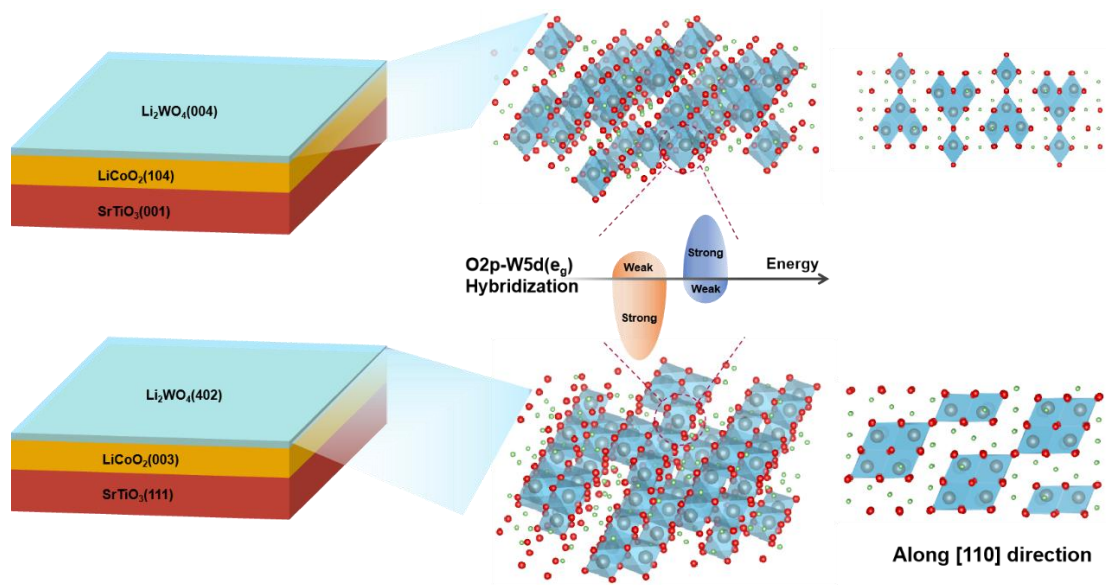

Figure 1. Schematic diagram of orientation-dependent electronic structure of Li$_2$WO$_4$.

## 2. Results and discussion

LiCoO$_2$ thin-films were grown on STO (001) and STO (111) single crystal substrates by pulsed laser deposition (PLD). During deposition process, the substrates were heated to 600°C and the growth oxygen pressure (PO$_2$) was fixed at 10 mTorr. After the synthesis process of the LiCoO$_2$ cathode seed layer, the substrate temperature was kept at 600°C while the PO$_2$ was increased to 100 mTorr for the deposition of the WO$_3$ film. The out-diffused Li ions from the LiCoO$_2$ cathode seed layers can react with the

$WO_3$ on the surface to form $Li_2WO_4$ during the epitaxial growth. In this case, the LWO film will turn out to be in a spinel structure under a tetragonal ($a = b = 11.94$ Å, $c = 8.41$ Å), space group $I4_1/amd$. Four $WO_6$ octahedra results in a $W_4O_{16}$ group by sharing octahedral edges and the Li atoms are filled into the oxygen octahedral and tetrahedral voids around the $WO_6$ octahedra[19,20]. A single layer of $WO_3$ was also grown on the STO substrates under the same growth conditions for the purpose of comparison[18].

The X-ray diffraction (XRD) pattern of $WO_3$/STO(001), LWO/LCO/STO(001) and LWO/LCO/STO(111) are shown in Figure 2a where the $WO_3$/STO(001) sample displays a single set of (001) film peaks in addition to the STO substrate peaks (denoted by asterisks) and the result is consistent with previous reports[21,22]. XRD analysis of the LWO/LCO/STO(001) sample reveals a distinct single Bragg peak, notably different from that of $WO_3$, with the exception of the LCO(104) diffraction peak. This distinctive peak originates from the LWO layer and it aligns precisely with the anticipated LWO(004) peak. Similarly, the XRD pattern of the LWO/LCO/STO(111) sample exhibits a characteristic diffraction peak at ~36.8° which corresponds to LWO(402). This observation strongly suggests that the crystal structure of LWO remains consistent across both samples, thereby emphasizing the robustness and uniformity of the LWO layer in this epitaxial system.

Atomic force microscopy (AFM) is also performed for each film surface. As shown in Figures 2b and c, the LWO films show different morphologies. When deposited on LCO/STO(001), LWO mainly nucleates as elongated nanorods, but a small number of nano-islands have also been observed. Meanwhile, when deposited on LCO/STO(111), the LWO layer consists of isosceles triangle shaped islands. According to previous reports, the difference in the LWO film morphologies may be due to defects in both LCO orientations where twin boundaries in LCO/STO(001) act as anchoring sites for nucleation, while the antiphase boundaries in LCO/STO(111) serve as channels for Li diffusion.

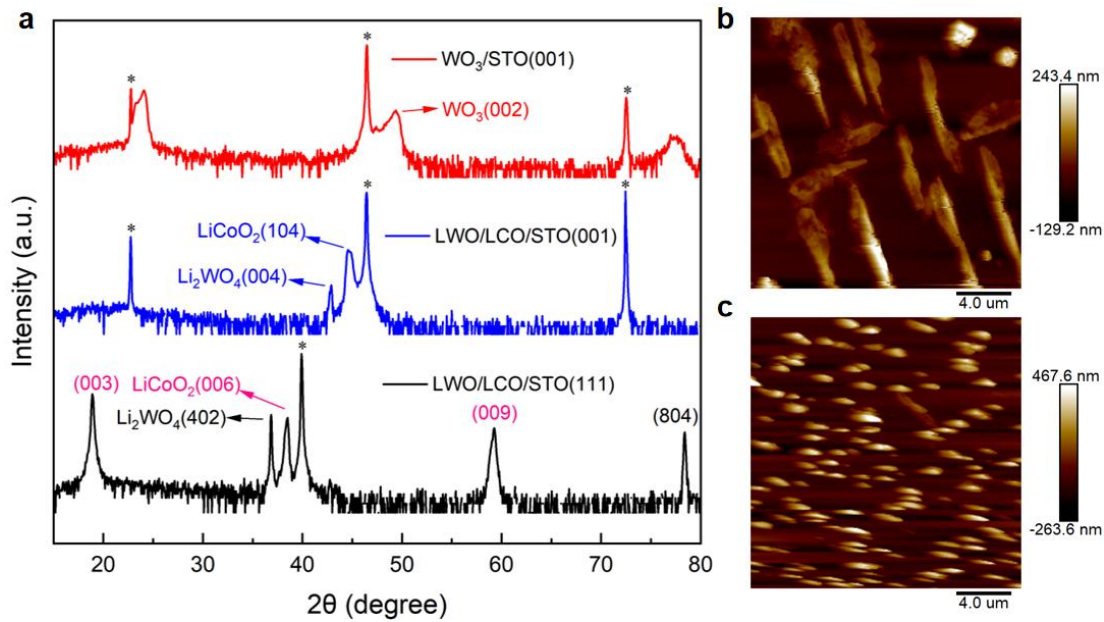

Figure 2. (a) X-ray diffraction θ−2θ scans of $WO_3$/STO(001), LWO/LCO/STO(001), and LWO/LCO/STO(111). AFM images of (b) LWO/LCO/STO(001) and (c) LWO/LCO/STO(111).

The XPS results of LWO/LCO/STO(001) and LWO/LCO/STO(111) have also been obtained and fitted. As shown in Figure 3, there are no obvious differences in the chemical bonds belonging to the O1s and W4f peaks between both samples. This implies that the growth process with different orientations does not alter the valence state of O and W components in LWO and LCO layer. Figure 3a shows that there are two peaks related to the valence state of the O-atoms in both samples. The distinct 530.7 eV peak is related to Co-O bonds[23] while the peak at 532.4 eV is related to W-O bonds[7]. Figure 3b exhibits the valence state of element W for both samples. It is clear that two main peaks located at ~35.7 and ~37.8 eV are related to the $4f_{5/2}$ and $4f_{7/2}$ of $W^{6+}$[14,22,24]. The analyses of the XPS data indicates that the LWO layer has been successfully synthesized on the LCO layer and retained the single valence state of W atom. Figure S1 also shows the survey XPS scan, which clearly shows that there is Co signal from LCO covered by LWO. This also indicates the incomplete coverage of LWO films.

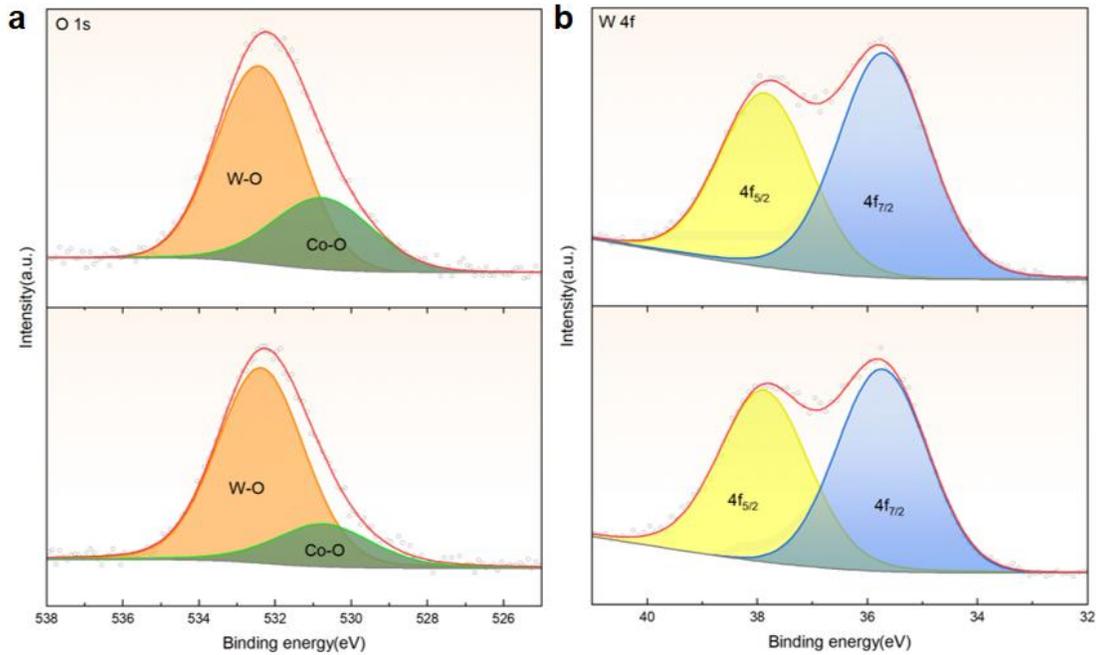

Figure 3. (a) O1s XPS of LWO/LCO/STO(001) (top) and LWO/LCO/STO(111) (bottom). (b) W4f XPS of LWO/LCO/STO(001) (top) and LWO/LCO/STO(111) (bottom)

Polarization-dependent XAS measurements were then performed where Figure 4a displays the modes of measurements at both grazing and normal angles of incidence. At grazing incidence, the E-vector of linear polarization X-ray is parallel to the out-of-plane c-axis (E//c) where the out-plane states are measured. As for measurements at normal incidence, the E-vector of linear polarization X-ray is parallel to the in-plane axes of the samples (E//ab). Hence, it registers principally the in-plane states. O K-edge measurements are carried out where it takes advantage of the transition from the O1s core level to O2p states which hybridize with the metallic d orbitals. This allows us to probe the empty states and the electronic structures of the LWO film.

Figures 4b, c display the polarization-dependent O K-edge XAS spectra of LWO/LCO/STO(001) and LWO/LCO/STO(111) respectively. The main spectral features are labelled peaks A, B, and C. As revealed by the AFM measurements (Figures 2b, c), the LCO layer is not totally covered by LWO films so that both LCO and LWO contribute to the intensity of spectra. However, comparing the results with the bare LCO/STO sample (Figure 4d), the results are consistent with previous studies[25–28], where some characteristic peaks of LWO would appear. The first principle feature labelled A (~530.5 eV) is attributed to an overlap of the O2p-Co3d($e_g$) hybridization[27,29] and the O2p-W5d($t_{2g}$) hybridization[30]. The energy positions of the two components are too close to distinguish. However, in comparison with bare LCO/STO, peak A shifts to a slightly higher photon energy and this suggests the addition of the LWO component[13]. In the spectral region of ~532-539 eV, peaks B (~532.4 eV) and C (~533.5 eV) correspond to the O2p-W5d($e_g$) hybridization[30]. The higher energy spectral region of ~536-545 eV are components from the Co4sp-O2p[31] and W6sp-O2p orbital hybridizations[30]. As for LWO/LCO/STO(001), the intensity of peak B considerably weakened and is barely visible. As for the case of LWO/LCO/STO(111), it follows the opposite trend where the intensity of peak C is significantly weaker than that of peak B and is barely noticeable. Hence, it can be deduced that while the O2p orbitals is mainly hybridized with higher energy level W5d($e_g$) orbitals for LWO/LCO/STO(001), the O2p orbitals are mainly hybridized at a lower energy with the W5d($e_g$) orbitals for LWO/LCO/STO(111).

In addition, the two growth orientations of LWO display different degrees of polarization dependence. As shown in Figure 4b, polarization dependence can be observed in the photon energy regions where features A, B and C are located for

LWO/LCO/STO(001). At higher photon energy, the polarization dependence is no longer noticeable. However, significant differences can be observed for LWO/LCO/STO(111) as showed in Figure 4c which indicates a large anisotropy between the in-plane and out-of-plane components. The experimental findings suggest that the electronic structure of crystalline LWO, especially the unoccupied O2p-W5d($e_g$) hybridized state, is strongly related to growth orientation which leads to different surface morphologic structure.

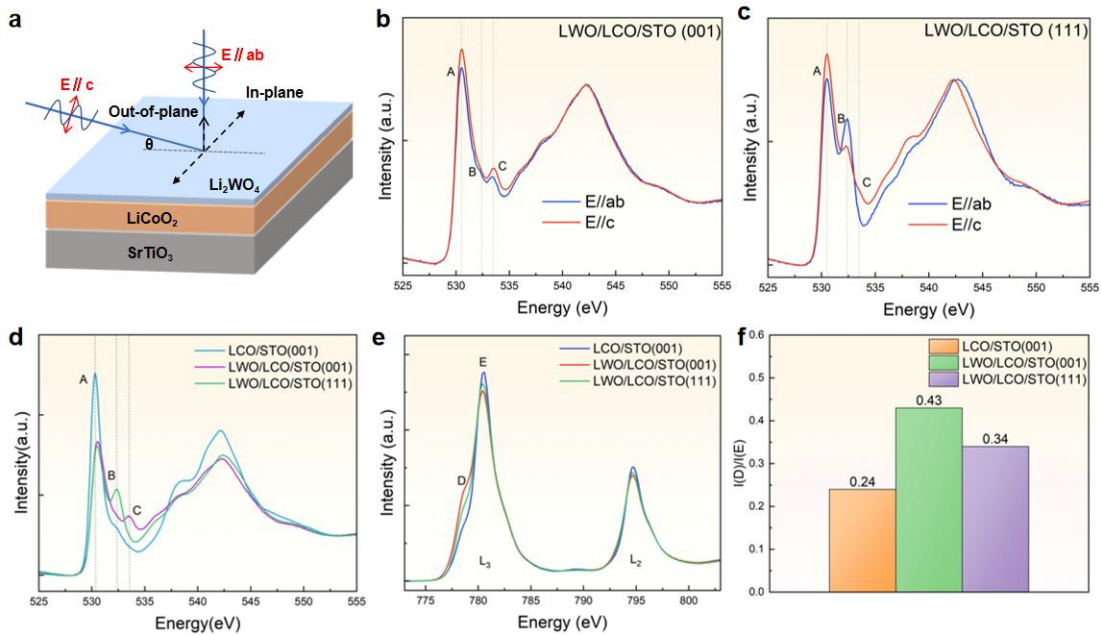

Figure 4. (a) Schematic illustrating the XAS experimental geometry, where *E* denotes the polarization of incoming photons and θ indicates the angle of incidence to the sample surface. O K-edge absorption spectra measured at normal (*E//ab*) and grazing (*E//c*) incidence for (b) LWO/LCO/STO(001) and (c) LWO/LCO/STO(111). The vertical dashed lines are drawn to guide the eyes. (d) O K-edge absorption spectra of three samples, with the average of *E//ab* and *E//c*. (e) Average Co $L_{2,3}$-edge absorption spectra of the three samples, with the average of *E//ab* and *E//c*. (f) The intensity ratio of peaks D and E in (e).

It should be noted that the growth of LWO is accompanied by Li-ion deintercalation in the LCO layer. During the delithiation process, charge is removed from the valence band and some empty states are generated. Figure 4e shows the Co L-edge XAS spectra of the LiCoO$_2$ system before and after the LWO layer is grown. The line shape is very similar to the LCO/STO spectrum where two main peaks (~794.8 eV and ~780.5eV) corresponding to the $L_2$- and $L_3$-edges are observed which can be attributed to transitions of Co2$p_{1/2}$ and Co2$p_{3/2}$ core electrons to the unoccupied 3d($e_g$) orbitals and they are highly hybridized with the O2p orbitals[32]. Nevertheless, a pre-peak appears with the growth of the LWO layer (labeled as D, at ~1.8 eV below the main peak E) and it can be attributed to the formation of Co$^{4+}$ ions[25]. We note that the intensity ratio between peak D and E is significantly different, thereby allowing us to compare the ratio of Co$^{4+}$ to Co$^{3+}$. This in turn will allow us to make a qualitative analysis of the degree of delithiation[33]. As shown in Figure 4f, for samples with STO(001)-oriented substrate, the intensity ratio I(D)/I(E) increases from 0.24 to 0.43 upon the introduction of the LWO layer. As for the LWO synthesized on the STO(111)-oriented substrate, the ratio is 0.34, smaller than the former but still larger than that of the bare LCO/STO. These results are consistent with previous reports concerning the channels for Li diffusion and defects in LCO[18].

To further illustrate how O2p-W5d orbital hybridization is manifested in the O K-edge XAS, we perform first principles calculations based on density functional theory method. By modelling the Li$_2$WO$_4$ lattice (Figure 5a), the density of states (DOS) of this structure could be properly elucidated (Figure 5b). More information about the distorted structure of WO$_6$ octahedron could be found in Figure S2. The calculated band gap of Li$_2$WO$_4$ is about 3.3 eV. The valence band (bandwidth of about 5.5 eV) is mainly composed of O2p states with some contribution from W5d states. In particular, there is a hybridization between the lower energy

O2p and W5d states, ranging approximately between -5.7 eV to -2.6 eV. The conduction band is mainly composed of W5d states with some contribution from O2p orbitals. Figure 5c and Figure S3 show the projected density of states (PDOS) of the W5d and O2p states separately. Since each W atom is surrounded by six O atoms in an octahedral coordination, the W5d states split into the $t_{2g}$ ($d_{xy}$, $d_{xz}$ and $d_{yz}$) and $e_g$ ($d_{z^2}$ and $d_{x^2-y^2}$) states attributed to crystal field theory. The central energy of the $t_{2g}$ band is about 4.2eV. Meanwhile, the $e_g$ states split into two separate parts at 6.7 and 8.1 eV, respectively. Principally, the results of these calculation studies show clear consistency with characteristic peaks A, B and C from O K-edge XAS spectra.

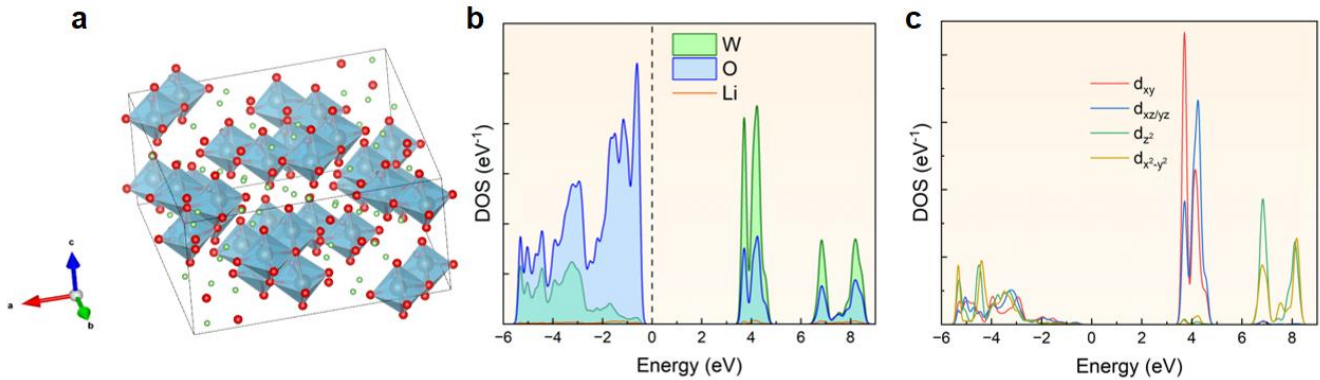

Figure 5. (a) Unit cell of the tetragonal $Li_2WO_4$ lattice. (b) Density functional theory calculations of the DOS of $Li_2WO_4$. The vertical dashed line indicates the zero energy point. (c) Projected density of states (PDOS) of the W5d states where the $d_{xz}$ and $d_{yz}$ orbitals are shown to be overlapping.

## 3. Conclusion

In summary, tetragonal structured LWO films have been synthesized on LCO layer with two orientations by PLD, and the electronic structure was investigated by polarization-dependent XAS. The experimental spectrum showed that different substrate orientations can tune the electronic structures of single-crystal LWO. The main O2p-W5d($e_g$) orbitals hybrid states of LWO/LCO/STO(001) is located at a higher energy level with a weaker in/out-of-plane anisotropy than that of LWO/LCO/STO(111) systems. The phenomena observed in this study may be attributed to the diffusion of Li ions, where different diffusion paths and $WO_6$ octahedral distortion have significant influence. These findings may assist in understanding how Li ions transport in LWO electrolyte and designing LCO cathode with better performance for all-solid-state LIBs.

## 4. Experimental

4.1. Sample preparation

$LiCoO_2$ films were grown on STO(001) and STO(111) single crystal substrates by pulsed laser deposition (PLD)[34]. Before growth, the STO(001) substrates were chemically etched in buffered HF acid to achieve $TiO_2$ termination and annealed in air at 1000 °C air for 1 hour to obtain a well-defined stepped surface structure. The UV laser pulse ($\lambda$ = 248 nm) has an energy density of ~2 J/cm2 and a repetition rate of 2 Hz. During deposition, the substrates were heated to 600°C with a growing oxygen pressure ($PO_2$) of 10 mTorr. After the growth of $LiCoO_2$ cathode seed layer, $WO_3$ films with a nominal thickness of ~10 to 30 nm were obtained by maintaining the substrates temperature at 600 °C and increasing the $PO_2$ to 100 mTorr. Here the repetition rate was reduced to 1hz. After growth, samples were cooled to room temperature at 100 mTorr.

4.2. Structural characterization

AFM (Bruker Dimension Edge) was used for the detection of surface morphology. High-resolution X-ray diffraction (XRD) measurements were performed at the X-ray Diffraction and Development (XDD) beamline at Singapore Synchrotron Light Source (SSLS) with an X-ray wavelength of $\lambda$ = 1.5404 Å.

4.3. XPS and XAS measurement

High-resolution XPS using a monochromatic Al-$K_\alpha$ X-ray source was carried out at normal emission with a VG/Scienta R3000 electron energy analyzer. Because of the sample load and the need to use a low-energy (1 eV) electronic flood gun, the spectrum of core energy levels was shifted to align the O1s peaks at 530 eV. The X-ray absorption spectroscopy (XAS) measurements were performed using linearly polarized X-ray at the Surface, Interface, and Nanostructure Science (SINS) beamline at SSLS. A total electron yield (TEY) detection method is used during the measurements. The incidence angle of X-rays refers to the sample surface was varied by rotating the polar angle of the sample. The spectra were measured at the incident angle of 20º ($E // c$) and 90º ($E // ab$). The spectra were normalized to the integrated intensity at the tail of the spectra after subtracting an energy-independent background.

4.4. First-principles calculations

First-principles calculations based on density functional theory (DFT) were undertaken by using Vienna ab initio simulation package (VASP)[35], adopting the projector-augmented wave method[36]. The exchange-correlation functional based on the generalized gradient approximation (GGA) in the form of the Perdew-Burke-Ernzerhof (PBE) functional[37], was used for both structural optimizations and electronic structure calculations. The energy cutoff was set at 750 eV. The density of k-point grids in the Brillouin zone is $7 \times 5 \times 5$. The convergence criteria for the structural relaxation was that the force and total energy were set to 0.001 eV/Å and $10^{-6}$ eV, respectively.

**Declaration of competing interest**

The authors declare no competing financial interest.

**Acknowledgement**

This work was supported by National Natural Science Foundation of China (Grant Nos.52172271, 12374378, 52307026), the National Key R&D Program of China (Grant No. 2022YFE03150200), Shanghai Science and Technology Innovation Program (Grant No. 22511100200, 23511101600). C.S.T acknowledges the support from the NUS Emerging Scientist Fellowship. The authors would like to acknowledge the Singapore Synchrotron Light Source for providing the facility necessary for conducting the research. The Laboratory is a National Research Infrastructure under the National Research Foundation, Singapore. Any opinions, findings, and conclusions or recommendations expressed in this material are those of the author(s) and do not reflect the views of National Research Foundation, Singapore.

**Supplementary materials**

**References**

[1] K Mizushima, P.C. Jones, P.J. Wiseman, J.B, Goodenough, Mater. Res. Bull. 15 (1980) 783–789.

[2] M.S. Whittingham, Chem. Rev. 104 (2004) 4271–4302.

[3] J.B. Goodenough, Y. Kim, Chem. Mater. 22 (2010) 587–603.

[4] Y. Lyu, X. Wu, K. Wang, Z. Feng, T. Cheng, Y. Liu, M. Wang, R. Chen, L. Xu, J. Zhou, Y. Lu, B. Guo, Adv. Energy Mater. 11 (2021) 2000982.

[5] J. Li, C. Lin, M. Weng, Y. Qiu, P. Chen, K. Yang, W. Huang, Y. Hong, J. Li, M. Zhang, C. Dong, W. Zhao, Z. Xu, X. Wang, K. Xu, J. Sun, F. Pan, Nat. Nanotechnol. 16 (2021) 599–605.

[6] Y. Zhu, N. Zhang, L. Zhao, J. Xu, Z. Liu, Y. Liu, J. Wu, F. Ding, J. Alloys Compd. 811 (2019) 152023.

[7] F. Meng, H. Guo, Z. Wang, G. Yan, X. Li, J. Alloys Compd. 790 (2019) 421–432.

[8] Z. Zhu, D. Yu, Z. Shi, R. Gao, X. Xiao, I. Waluyo, M. Ge, Y. Dong, W. Xue, G. Xu, W.-K. Lee, A. Hunt, J. Li, Energy Environ. Sci. 13 (2020) 1865–1878.

[9] J. Cabana, B.J. Kwon, L. Hu, Acc. Chem. Res. 51 (2018) 299–308.

[10] Y. Xu, E. Hu, K. Zhang, X. Wang, V. Borzenets, Z. Sun, P. Pianetta, X. Yu, Y. Liu, X.-Q. Yang, H. Li, ACS Energy Lett. 2 (2017) 1240–1245.

[11] T. Yoshimura, M. Watanabe, Y. Koike, K. Kiyota, M. Tanaka, Jpn. J. Appl. Phys. 22 (1983) 152.

[12] T. Hayashi, T. Miyazaki, Y. Matsuda, N. Kuwata, M. Saruwatari, Y. Furuichi, K. Kurihara, R. Kuzuo, J. Kawamura, J. Power Sources 305 (2016) 46–53.

[13] T. Hayashi, J. Okada, E. Toda, R. Kuzuo, Y. Matsuda, N. Kuwata, J. Kawamura, J. Power Sources 285 (2015) 559–567.
[14] Z. Sun, Y. Lai, N. Lv, Y. Hu, B. Li, S. Jing, L. Jiang, M. Jia, J. Li, S. Chen, F. Liu, Adv. Mater. Interfaces 8 (2021) 2100624.
[15] C. Qin, L. Wang, P. Yan, Y. Du, M. Sui, Chin. Phys. Lett. 38 (2021) 068202.
[16] M. Hirayama, N. Sonoyama, T. Abe, M. Minoura, M. Ito, D. Mori, A. Yamada, R. Kanno, T. Terashima, M. Takano, K. Tamura, J. Mizuki, J. Power Sources 168 (2007) 493–500.
[17] W. Samarakoon, J. Hu, M. Song, M. Bowden, N. Lahiri, J. Liu, L. Wang, T. Droubay, K. Koirala, H. Zhou, Z. Feng, J. Tao, Y. Du, J. Phys. Chem. C 126 (2022) 15882–15890.
[18] L. Wang, Z. Yang, W.S. Samarakoon, Y. Zhou, M.E. Bowden, H. Zhou, J. Tao, Z. Zhu, N. Lahiri, T.C. Droubay, Z. Lebens-Higgins, X. Yin, C.S. Tang, Z. Feng, L.F.J. Piper, A.T.S. Wee, S.A. Chambers, Y. Du, Nano Lett. 22 (2022) 5530–5537.
[19] H. Horiuchi, N. Morimoto, S. Yamaoka, J. Solid State Chem. 30 (1979) 129–135.
[20] C.W.F.T. Pistorius, J. Solid State Chem. 13 (1975) 325–329.
[21] Y. Du, M. Gu, T. Varga, C. Wang, M.E. Bowden, S.A. Chambers, ACS Appl. Mater. Interfaces 6 (2014) 14253–14258.
[22] H. Kalhori, S.B. Porter, A.S. Esmaeily, M. Coey, M. Ranjbar, H. Salamati, Appl. Surf. Sci. 390 (2016) 43–49.
[23] X. Zheng, Y. Chen, X. Zheng, G. Zhao, K. Rui, P. Li, X. Xu, Z. Cheng, S.X. Dou, W. Sun, Adv. Energy Mater. 9 (2019) 1803482.
[24] S. Hashimoto, H. Matsuoka, H. Kagechika, M. Susa, K.S. Goto, J. Electrochem. Soc. 137 (1990) 1300–1304.
[25] E. Salagre, P. Segovia, M.Á. González-Barrio, M. Jugovac, P. Moras, I. Pis, F. Bondino, J. Pearson, R.S. Wang, I. Takeuchi, E.J. Fuller, A.A. Talin, A. Mascaraque, E.G. Michel, ACS Appl. Mater. Interfaces 15 (2023) 36224–36232.
[26] J. Van Elp, J.L. Wieland, H. Eskes, P. Kuiper, G.A. Sawatzky, F.M.F. De Groot, T.S. Turner, Phys. Rev. B 44 (1991) 6090–6103.
[27] D. Ensling, A. Thissen, S. Laubach, P.C. Schmidt, W. Jaegermann, Phys. Rev. B 82 (2010) 195431.
[28] W.-S. Yoon, K.-B. Kim, M.-G. Kim, M.-K. Lee, H.-J. Shin, J.-M. Lee, J.-S. Lee, C.-H. Yo, J. Phys. Chem. B 106 (2002) 2526–2532.
[29] W. Zhang, E. Hosono, D. Asakura, H. Yuzawa, T. Ohigashi, M. Kobayashi, H. Kiuchi, Y. Harada, Sci. Rep. 13 (2023) 4639.
[30] M.B. Johansson, P.T. Kristiansen, L. Duda, G.A. Niklasson, L. Österlund, J. Phys. Condens. Matter 28 (2016) 475802.
[31] A. Juhin, F. De Groot, G. Vankó, M. Calandra, C. Brouder, Phys. Rev. B 81 (2010) 115115.
[32] T. Mizokawa, Y. Wakisaka, T. Sudayama, C. Iwai, K. Miyoshi, J. Takeuchi, H. Wadati, D.G. Hawthorn, T.Z. Regier, G.A. Sawatzky, Phys. Rev. Lett. 111 (2013) 056404.
[33] T.K. Sham, J. Chem. Phys. 79 (1983) 1116–1121.
[34] Z. Yang, P. Ong, Y. He, L. Wang, M.E. Bowden, W. Xu, T.C. Droubay, C. Wang, P.V. Sushko, Y. Du, Small 14 (2018) 1803108.
[35] G. Kresse, J. Furthmüller, Phys. Rev. B 54 (1996) 11169–11186.
[36] P.E. Blöchl, Phys. Rev. B 50 (1994) 17953–17979.
[37] J.P. Perdew, K. Burke, M. Ernzerhof, Phys. Rev. Lett. 77 (1996) 3865–3868.